\documentclass[aps,prl,twocolumn,showpacs,superscriptaddress]{revtex4-1}

\usepackage{graphicx,charter,helvet,sansmath,color,cancel}
\usepackage[normalem]{ulem}
\usepackage{natbib}

\begin{document}

\title{Resolving interfacial charge transfer in titanate superlattices using resonant X-ray reflectometry}

\author{R. F. Need}
 \email{ryan.need@nist.gov}
 \affiliation{NIST Center for Neutron Research, National Institute of Standards and Technology, Gaithersburg, Maryland 20899, USA}
 
\author{P. B. Marshall}
 \affiliation{Materials Department, University of California, Santa Barbara, California 93106, USA}

\author{E. Weschke}
 \affiliation{Institute for Quantum Phenomena in Novel Materials, \\ Helmholtz Zentrum Berlin, Adlershof, Germany}
 
 \author{A. J. Grutter}
  \affiliation{NIST Center for Neutron Research, National Institute of Standards and Technology, Gaithersburg, Maryland 20899, USA}
 
 \author{D. A. Gilbert}
  \affiliation{NIST Center for Neutron Research, National Institute of Standards and Technology, Gaithersburg, Maryland 20899, USA}

 \author{E. Arenholz}
  \affiliation{Advanced Light Source, Lawrence Berkeley National Laboratory, Berkeley, California 94720, USA}
  
 \author{P. Shafer}
  \affiliation{Advanced Light Source, Lawrence Berkeley National Laboratory, Berkeley, California 94720, USA}

\author{S. Stemmer}
 \affiliation{Materials Department, University of California, Santa Barbara, California 93106, USA}

\author{S. D. Wilson}
 \affiliation{Materials Department, University of California, Santa Barbara, California 93106, USA}

\begin{abstract}
Charge transfer in oxide heterostructures can be tuned to promote emergent interfacial states, and accordingly, has been the subject of intense study in recent years. However, accessing the physics at these interfaces, which are often buried deep below the sample surface, remains difficult. Addressing this challenge requires techniques capable of measuring the local electronic structure with high-resolution depth dependence. Here, we have used linearly-polarized resonant X-ray reflectometry (RXR) as a means to visualize charge transfer in oxide superlattices with atomic layer precision. From our RXR measurements, we extract valence depth profiles of SmTiO$_3$ (SmTO)/SrTiO$_3$ (STO) heterostructures with STO quantum wells varying in thickness from 5 SrO planes down to a single, atomically thin SrO plane. At the polar-nonpolar SmTO/STO interface, an electrostatic discontinuity leads to approximately half an electron per areal unit cell transferred from the interfacial SmO layer into the neighboring STO quantum well. We observe this charge transfer as a suppression of the t$_{2g}$ absorption peaks that minimizes contrast with the neighboring SmTO layers at those energies and leads to a pronounced absence of superlattice peaks in the reflectivity data. Our results demonstrate the sensitivity of RXR to electronic reconstruction at the atomic scale, and establish RXR as a powerful means of characterizing charge transfer at buried oxide interfaces.
\end{abstract}

\maketitle

Controlling charge transfer at heterointerfaces to create high-density two-dimensional electron systems (2DES) has been one of the most important developments in electronic materials over the last four decades. The topic traces its origins back to work in III-V compounds, where high mobility GaAs/AlGaAs interfaces enabled new transistor designs and led to the discovery of the fractional quantum Hall effect \cite{mimura1980new,tsui1982two}. More recently, improvements in the synthesis of complex oxide films have made it possible to study oxide heterostructures with atomically sharp interfaces \cite{chambers2010epitaxial,boschker2017quantum}, many of which also display interfacial charge transfer \cite{mannhart2010oxide,Stemmer20142DEGs}. However, unlike their III-V analogues, the valence electrons in complex oxide heterostructures occupy strongly correlated \textit{d} orbitals leading to a host of emergent phenomenon at their interfaces \cite{Hwang2012Emergent}. From superconductivity to magnetism, the unique ground states found at oxide interfaces hold incredible promise to develop new classes of electronics and expand our understanding of correlated electron behavior in materials.

Moving from the promise of these ideas to their realization requires the ability to finely tune charge transfer at interfaces by varying parameters like composition, layer thicknesses, and intermixing at interfaces. This path forward also requires the ability to precisely characterize the resulting interfacial electronic structure. Transport and Hall measurements are invaluable techniques for understanding the collective behavior of 2DES \cite{tsukazaki2010observation,caviglia2008electric,Mikheev2015Separation}, but they are indirect and probe only the itinerant electrons. Other techniques offer direct access to the local electronic structure, but are either restricted to surface of the sample by small electron escape depths (e.g. photoemission spectroscopy \cite{berner2013direct}) or probe only a small areal fraction of the sample (e.g. cross sectional scanning tunneling spectroscopy (STS) \cite{kourkoutis2013visualizing,huang2018atomically}, scanning transmission microscopy with electron energy loss spectroscopy (STEM-EELS) \cite{ohtomo2002artificial,muller2009structure}).

Resonant X-ray reflectometry (RXR) presents a different means of resolving charge transfer at buried interfaces in an oxide superlattice. RXR combines the atomic layer depth resolution of hard X-ray reflectometry with the sensitivity to valence electrons inherent to X-ray absorption spectroscopy (XAS) \cite{macke2014element,zwiebler2015electronic}. The result is a probe capable of measuring the electronic, magnetic, and orbital properties of buried interfaces in complex heterostructure geometries over macroscopic sample areas \cite{Benckiser2011Orbital,chen2015extreme,hamann2017site}.

To date, only a few attempts to investigate interfacial charge transfer using RXR have been conducted. An early example used energy scans of a nominally forbidden superlattice reflectivity peak to study electronic reconstruction at the LaMnO$_3$/SrMnO$_3$ interface, but did not attempt to refine the spatial distribution of that reconstruction \cite{smadici2007electronic}.  More recently, Hamann-Borrero \textit{et al.}  reported valence depth profiles in monolithic LaCoO$_3$ films, where they observed electronic reconstruction in the two unit cells near an uncapped surface \cite{hamann2016valence}. 

In this study, we examine SmTiO$_3$ (SmTO)/SrTiO$_3$ (STO) superlattices where the titanium valence state changes from, nominally, Ti$^{3+}$ in SmTO to Ti$^{4+}$ in STO. This broken symmetry creates an electrostatic discontinuity in the film that causes half an electron per areal unit cell to transfer from SmTO to STO, forming a high density (3 x 10$^{14}$ cm$^{-2}$) 2DES bound to the interface \cite{Stemmer2013Quantum}. Using RXR, we detect this charge transfer as a suppression of the STO t$_{2g}$ absorption peaks relative to a Ti$^{4+}$ standard, which indicates increased t$_{2g}$ orbital occupation. Unlike the previous LaCoO$_3$ work \cite{hamann2016valence}, both the chemical composition and transition metal valence state change across the interface in our samples. Importantly, our results demonstrate the ability to disentangle these two contrast mechanisms and directly resolve electronic reconstruction.  By comparing multiple potential models, we prove that different local Ti electronic environments are required within the SmTO and STO layers, and that charge transfer can be detected down to the length scale of a single atomic plane. Finally, these results are obtained despite the presence of unwanted surface oxidation layers that are difficult to avoid in many rare earth-containing oxide films.

Three samples were grown for these experiments using hybrid molecular beam epitaxy (MBE) to deposit films on (La$_{0.18}$Sr$_{0.82}$)(Al$_{0.59}$Ta$_{0.41})$O$_{3}$ (LSAT) (001) substrates \cite{Jalan2009Growth,Moetakef2013Growth}. A single 20 nm film of SmTO was used to provide a structural and chemical baseline for analyzing charge transfer in the accompanying superlattice samples. The two SmTO/STO superlattices had the general structure, LSAT/[X SmTO/Y STO]$_{4}$/X SmTO, where X and Y are the thicknesses of the SmTO and STO layers, respectively, quantified in unit cells of SmTO and number of SrO planes. The superlattice architectures (X:Y) 16:1 and 10:5 were chosen to test RXR's ability to resolve charge transfer in the atomic limit, while keeping the total film thickness below the critical thickness for pseudomorphic growth.

\begin{figure}
\centering \includegraphics[width=2.5in]{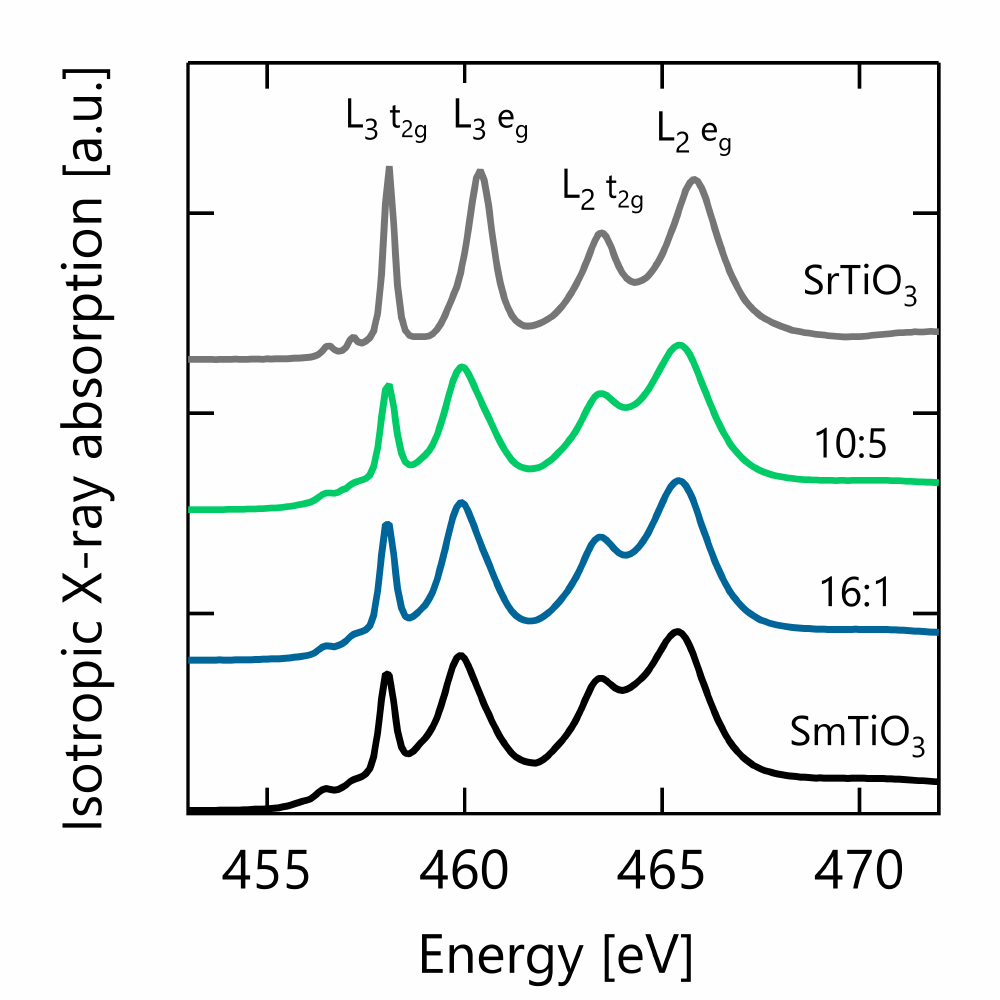}
\caption{Isotropic TEY X-ray absorption spectra for a SmTO control film, STO substrate, and two SmTO/STO superlattices measured at 300 K. The labels X:Y correspond to the SmTO:STO layer architectures of the superlattices as described in the main text. \label{fig:SmTORXR_2}}  
\end{figure}

Figure \ref{fig:SmTORXR_2} contains the polarization-averaged or isotropic X-ray absorption spectra for all three samples and a bare STO substrate measured in total electron yield (TEY) mode. This data was used to extract the complex anomalous scattering factors, $f^{\prime}$ and $f^{\prime}$$^{\prime}$, specific to the local Ti environment in our samples by scaling the measured XAS to tabulated values of the Ti X-ray scattering factors \cite{chantler2000detailed,SuppInfo}. These values contain the energy-dependent changes in scattering and absorption for Ti near its resonant energy transition, and are critically necessary input to properly refine the RXR data.

XAS from the STO substrate provided a control spectra for Ti$^{4+}$ in O$_{h}$ symmetry, and indeed the measured data matches well with previous reports \cite{DeGroot1992,lesne2014suppression}. Similarly, the SmTO film was meant to provide a control of the Ti$^{3+}$ spectra. However, as Fig. \ref{fig:SmTORXR_2} clearly shows, the SmTO film---and both superlattices---have spectra resembling Ti$^{4+}$. This does not indicate some unusual Sm valence, but rather results from the oxidation of the surface of rare earth transition metal oxide films \cite{macke2014element,zwiebler2015electronic,Need2016Interface}. Previous studies have shown surface oxidation in these materials extends 2 nm to 3 nm from the surface, which constitutes the majority of the           $\approx$ 3 nm electron sampling depth calculated for similar perovskites \cite{ruosi2014electron}. As a result of this surface oxidation, the scattering factors we extract from scaling the XAS data are not representative of the underlying, unoxidized layers of our film. However, we show that this difficulty can be overcome through careful refinement of the buried scattering factors using the depth-dependent reflectometry data.

\begin{figure}
\centering \includegraphics[width=2.5in]{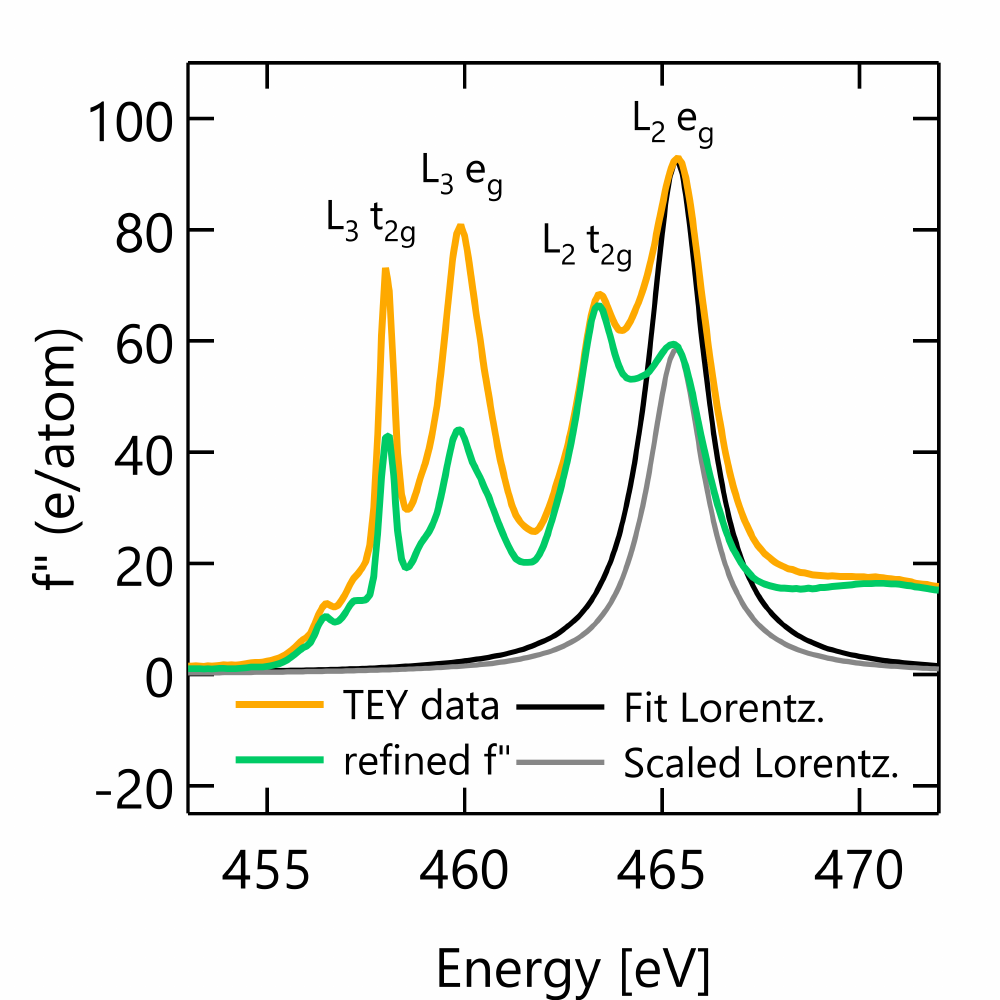}
\caption{Measured TEY (yellow) and refined (green) spectra of the imaginary anomalous scattering factor for the SmTO control film. The fit (black) and scaled (grey) Lorentzian peaks were used to refine the measured XAS profile as described in the main text. \label{fig:SmTORXR_3}}  
\end{figure}

To demonstrate the process of refining scattering factors in buried layers, we use the 20 nm SmTO control film as an example. As shown in Fig. \ref{fig:SmTORXR_3} for the L$_2$ e$_g$ peak, the resonant scattering factors of the buried Ti slab were refined by fitting each Ti resonance peak in the measured $f^{\prime}$$^{\prime}$ spectra (yellow) with a Lorentzian line shape (black) \cite{PhysRevB.82.235410}. This Lorentzian fit was then subtracted from the measured $f^{\prime}$$^{\prime}$ spectrum. Finally, the same Lorentzian was added back to the $f^{\prime}$$^{\prime}$ spectrum \textit{multiplied by a scaling parameter that was free to fit during during the reflectivity refinement} (grey), resulting in a refined scattering factor spectra (green). Note only at the four absorption peak positions where RXR data were collected is the scattering factor profile is accurately refined. Changes throughout the rest of the profile are due to the tails of the Lorentzian peaks used to modify the scattering factor spectrum. A major advantage of this Lorentzian fitting method is that it allows the physically-required Kramers-Kronig (KK) relation between  $f^{\prime}$ and  $f^{\prime}$$^{\prime}$ to be preserved. Specifically, the real part of the scattering factor,  $f^{\prime}$, is adjusted by the KK transform of the Lorentzian peak fit, multiplied by the same scaling parameter that adjusts the amplitude of the Lorentzian fit to the imaginary component, $f^{\prime}$$^{\prime}$. This procedure also adds only one free parameter per peak to the total number of refined variables keeping added computational expense to a minimum.

\begin{figure}
\includegraphics[width=2.5in]{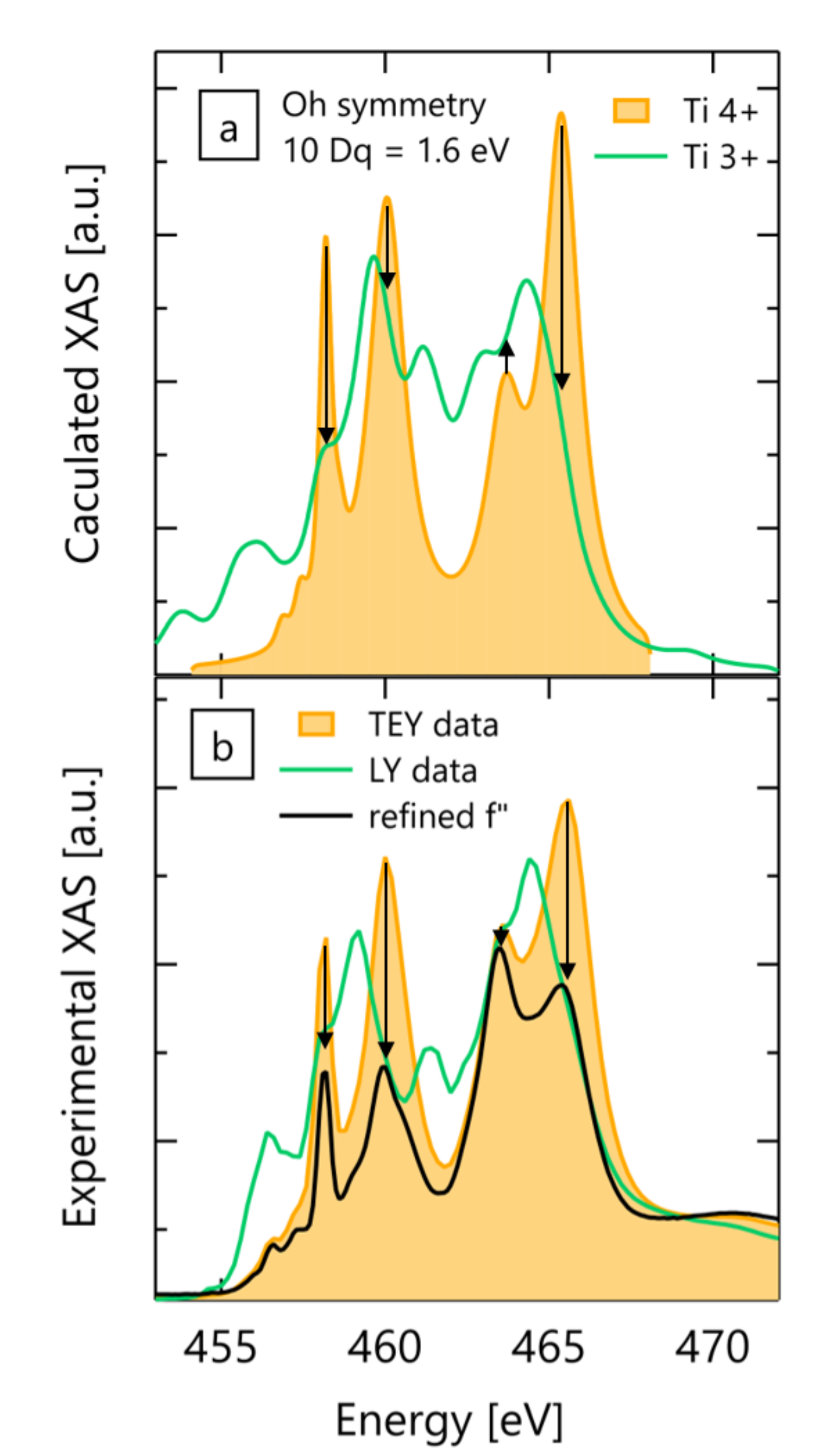}
\caption{Crystal field multiplet calculations (a) for Ti$^{4+}$ and Ti$^{3+}$ ions in O$_h$ symmetry made using the program CTM4XAS. X-ray absorption spectra (b) from the 20 nm SmTO control film measured in TEY (yellow) and LY (green) modes, which probe the oxidized surface and total film thickness respectively, are compared to the fit $f^{\prime}$$^{\prime}$ spectra (black). \label{fig:SmTORXR_4}}  
\end{figure}

While the refined $f^{\prime}$$^{\prime}$ spectrum in Fig. \ref{fig:SmTORXR_3} still slightly resembles Ti$^{4+}$, it must to some degree because of the limited number of discreet energies at which the spectrum was refined. Therefore, in order to understand and interpret the refined $f^{\prime}$$^{\prime}$ spectrum, we compare it to crystal field multiplet calculations and bulk sensitive luminescence yield (LY) data in Fig. \ref{fig:SmTORXR_4}(a) and \ref{fig:SmTORXR_4}(b), respectively. In both cases, we found a ``fingerprint'' pattern of peak adjustments that match that of the refined spectrum. More specifically, starting with a Ti$^{4+}$ spectrum (yellow) and moving towards a Ti$^{3+}$ pattern (green) requires substantial suppression of both L$_3$ peaks and the L$_2$ e$_g$ peak (1st, 2nd, and 4th) while the L$_{2}$ t$_{2g}$ shows a smaller and more ambiguous change. Close comparison of the different spectra shows that the experimentally measured XAS in Fig. \ref{fig:SmTORXR_4}(b) is a better match to the refined spectrum than the multiplet calculations. This is likely because the bulk sensitive LY experimental data captures slight deviations from octahedral symmetry of Ti$^{3+}$ ions in the SmTO film that were not included in the multiplet calculations.

The strong agreement between our refinements, theoretical calculations, and LY XAS data confirm that the refined $f^{\prime}$$^{\prime}$ spectrum in Fig. \ref{fig:SmTORXR_3} and \ref{fig:SmTORXR_4}(b) reflect the presence of unoxidized SmTO with Ti$^{3+}$ beneath our oxidized surface layer. Moreover, this result demonstrates that information about the local coordination and electronic state of buried layers can be extracted by refining the scattering factors from a relatively small density of reflectivity curves. Though such results should be corroborated with comparison to theory or alternative probes, this nonetheless has important implications for the analysis of buried layers in multilayer thin films that cannot accurately be measured by other means or where surface oxidation obscures its direct measurement.

\begin{figure}
\includegraphics[width=3in]{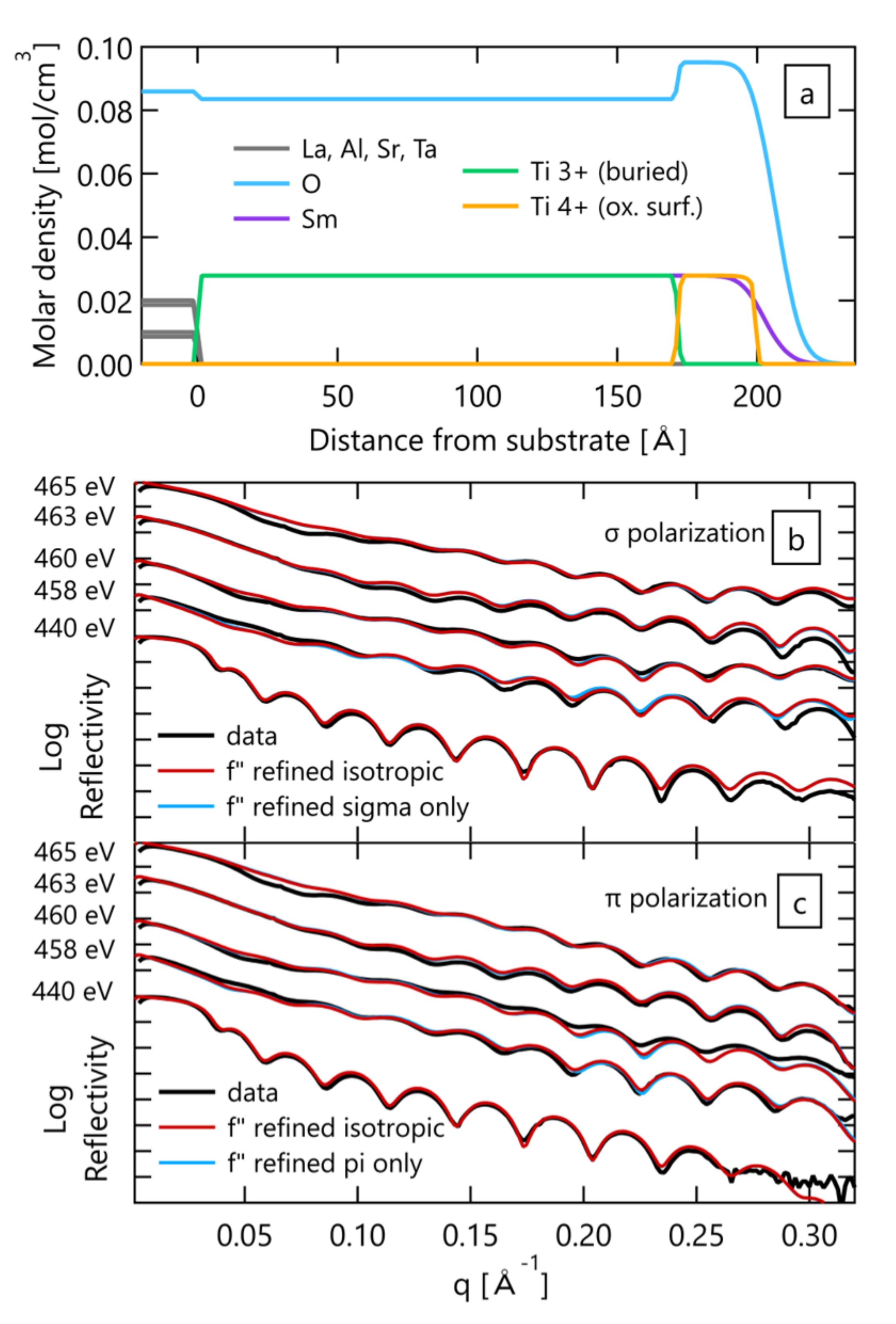}
\caption{Refined sample model (a) showing the various elemental profiles including the oxidized Ti surface layer. Sigma (b) and pi (c) polarization reflectivity data and fits from the refined model, comparing models with scattering factor spectra that were refined under isotropic and anisotropic assumptions. \label{fig:SmTORXR_5}}  
\end{figure}

Figure \ref{fig:SmTORXR_5}(a) displays the structural layer model of the 20 nm SmTO control film that was refined simultaneously with the scattering factor spectrum just discussed. The Ti depth profile of this sample consists of just two slabs---a thinner, oxidized surface layer and the thicker, unoxidized buried layer. The individual Ti layer thicknesses and roughnesses were not constrained and allowed to adjust freely. The atomic densities were fixed to their theoretical stoichiometric values with the exception of oxygen in the two topmost slabs (oxygen profile had three slabs). This assumption was deemed appropriate given the established presence of a stoichiometric growth window for both STO and SmTO grown by hybrid MBE \cite{Jalan2009Growthwindow, Moetakef2013Growth}.

From the refinement, we find that the thickness of the surface oxidation is 2.78 nm, similar to previous reports on related films. This is observed as an increase in the oxygen density profile near the surface, as well as the Ti$^{4+}$ valence state discussed above. The oxidized Ti has an abrupt top interface with 0.8 {\AA} rms roughness, considerably sharper than both the Sm and O profiles, which are 7.0 {\AA} and 5.9 {\AA}, respectively. The same sequence of layer termination and the relative roughness at the surface interface was found in all three samples through independent refinements. This repeatability indicates the surface profile here is not arbitrary, but a consequence of the growth process that may reflect the length of time various molecular species linger in the growth chamber or a preference for SmO surface termination.

The quality of the fits produced by this model is shown in Figs. \ref{fig:SmTORXR_5}(b) and \ref{fig:SmTORXR_5}(c) plotted against the 300 K reflectivity data, which constitutes just 10 of the 26 co-refined datasets for this sample. In fact, excellent agreement is achieved for all 26 datasets over an intensity range spanning seven orders of magnitude \cite{SuppInfo}. Also shown in panels (b) and (c) is the negligible change that occurs when the assumption of isotropic scattering factors (red) is removed, and unique scattering factors (blue) are refined to the sigma and pi polarization datasets for a fixed sample profile. This comparison is the same as comparing isotropic and tetragonal scattering matrices, or equivalently, reducing the Ti point group symmetry from O$_h$ to D$_{4h}$. Given the $\approx$1$\%$ average compressive in-plane strain from epitaxial growth on LSAT, and the suggestion of reduced symmetry from the LY XAS, one could reasonably assume a tetragonal scattering matrix would better capture the RXR data. However, the fit quality in Fig. \ref{fig:SmTORXR_5} shows that this reduced symmetry is not apparent in the data, and therefore, we restricted our analysis to the simpler picture of isotropic scattering factors.

Information gathered from fitting the 20 nm SmTO control film was used to better analyze the superlattice structures. Specifically, the topmost layer of each superlattice structure, which in both cases was $\geq$4 nm of SmTO, was broken up into two Ti slabs to account for a similar amount of surface oxidation as seen in the control film. For the buried SmTO layers, the scattering factors of the associated Ti slabs were refined in the same manner as before, using the Lorentzian peak scaling parameters from the control film as starting values. The buried STO quantum wells were treated as uniform, and the scattering factors for Ti in the STO quantum wells were refined in an analogous manner to the Ti in the SmTO layers. Finally, the interface roughnesses were constrained to be uniform for a given interface type (i.e. SmTO-to-STO vs. STO-to-SmTO).

\begin{figure*}[th]
\centering \includegraphics[width=7in]{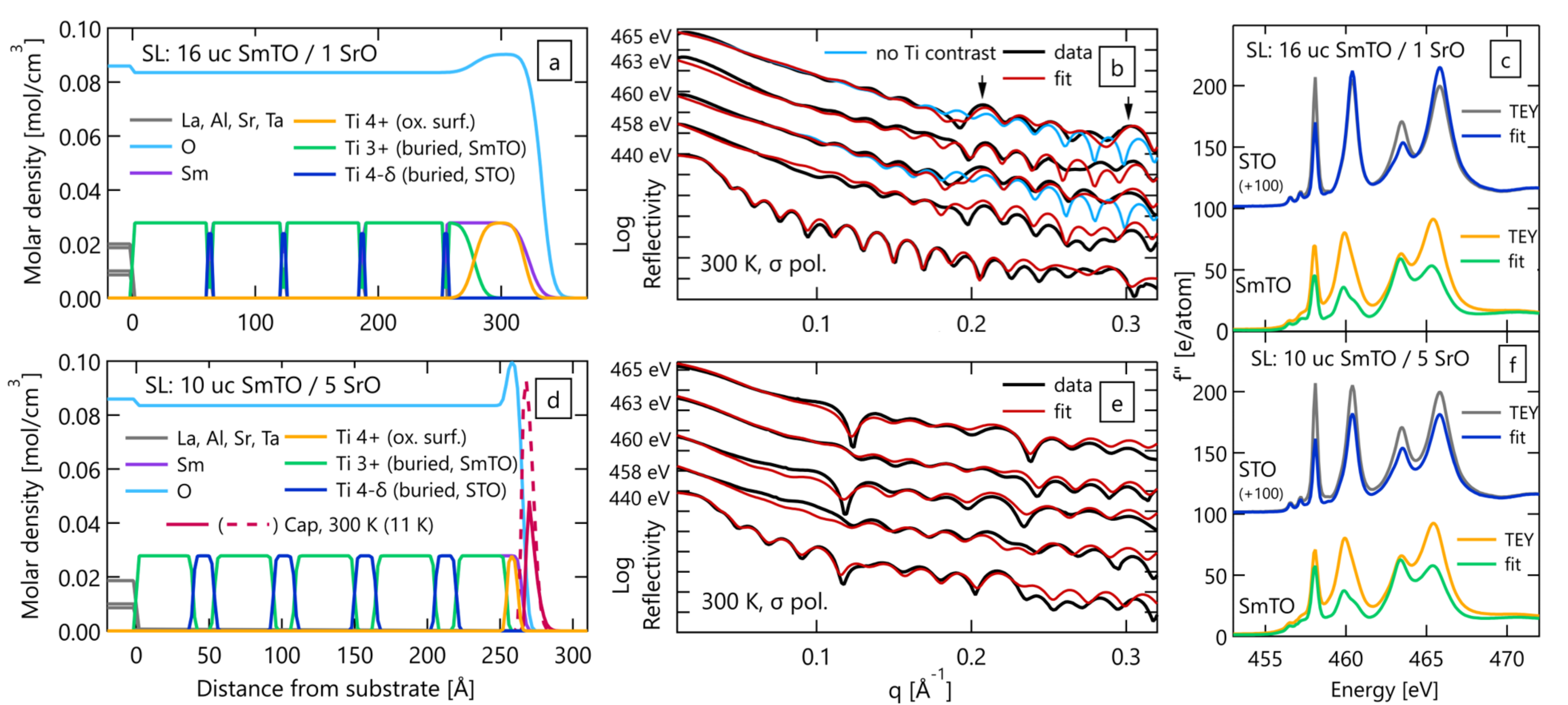}
\caption{Results of reflectivity refinements for the two superlattice samples. Sample layer models (a) and (d) show each sample's structure through the combination of multiple elemental depth profiles. Fits (red) to 300 K sigma reflectivity data (black) are plotted in panels (b) and (e). Also in panel (b) are the poor fits (light blue) from a comparative model without Ti contrast as described in the main text. Arrows point to superlattice peaks. Finally, the measured (grey and yellow) and refined (blue and green) $f^{\prime}$$^{\prime}$ spectra for the STO and SmTO layers are shown in panels (c) and (f) for the 16:1 and 10:5 superlattices, respectively. \label{fig:SmTORXR_6}}  
\end{figure*}

The resulting sample models of the SmTO/STO superlattices are shown in Fig. \ref{fig:SmTORXR_6}(a) and (d), along with the corresponding fits to the 300 K sigma polarization data (b) and (e). Considering first the sample profiles, in both cases we clearly observe the surface termination pattern described above, with oxidiation thicknesses varying between 0.90 nm and 4.10 nm. In the case of the 10:5 superlattice geometry, a carbon contamination layer, assocaited with non-resonant temperature-dependent reflectivity changes (see Supporting Information), is shown for both room and base temperatures. More importantly, the refined structures all contain an alternating pattern of Ti$^{3+}$/ Ti$^{4-\delta}$ slabs. For the 1 SrO superlattice, the STO layers thickness were refined to be 1.6 {\AA} thick. This value is slightly smaller than half a unit cell of STO \cite{Jalan2009Growth}, indicating these layers indeed correspond to a single SrO plane and the two TiO$_2$ planes on either side. To extract information about the charge transfer in these interfacial layers, we now turn to the the refined spectra for these Ti$^{3+}$ and Ti$^{4-\delta}$ layers and their interpretation

Figures \ref{fig:SmTORXR_6} (c) and (f) contain the measured and refined scattering factor profiles for Ti in the STO and SmTO layers for the 16:1 and 10:5 superlattices respectively. As expected, the $f^{\prime}$$^{\prime}$ spectra refined for the SmTO layers are nearly identical to the pristine Ti$^{3+}$ specta determined for the SmTO control film. For the STO layers, the refined Ti $f^{\prime}$$^{\prime}$ spectra also show striking similarities between samples, despite a variation in quantum well thickness from 1.6 {\AA} to 15.2 {\AA}. Specifically, the STO e$_g$ peaks shift only slightly relative to the TEY spectra, while the t$_{2g}$ peaks are more notably and uniformly suppressed. The retention of e$_g$ peak intensity, where the greatest suppression of the Ti$^{3+}$ profiles occurs, provides a clear indication that the Ti in the STO layers possesses a Ti valence state distinct from the neighboring SmTO.

When interpreting the refined STO $f^{\prime}$$^{\prime}$ spectra, it is important to recall that the XAS spectra is roughly proportional to the unoccupied density of states \cite{DeGroot2005Multiplet}. Peak suppression then indicates an \textit{increased} occupation of the orbital states associated with that peak energy. Given that the suppressed peaks at 458.1 eV and 463.4 eV are well known to correspond to the Ti t$_{2g}$ orbitals \cite{DeGroot1992}, our results suggest that these t$_{2g}$ orbitals are more heavily populated in the thin STO layers of the superlattices than the fully d$^0$ Ti$^{4+}$ ions in the STO substrate used as a control. This observation fits well with the polar discontinuity and interfacial charge transfer that are known to exist in this system. As the roughly half an electron per areal unit cell is donated from the nearest SmO plane to the interfacial TiO$_2$ plane, these electrons would be expected to populate the lowest available energy orbitals, which in this system are the STO Ti t$_{2g}$ orbitals. Indeed, this is precisely what has been observed in the case of LaAlO$_3$/SrTiO$_3$, which contains a similar polar discontinuity and band alignment \cite{Salluzzo2009,berner2013direct}. In other words, the t$_{2g}$ peak suppression in the refined STO spectra indicates the presence of charge transfer at the SmTO/STO interface, and is consistent with the expected formation of a 2DES confined within the STO layers.

The spatial charge distribution of the 2DES is contained in the depth profile of the refined STO Ti$^{4-\delta}$ spectra. For both the 1 SrO and 5 SrO STO wells, the 2DES is confined to the STO layer and appears homogeneous throughout the well. This can best be understood by considering the band structure and alignment of the SmTO/STO interface. In the case of a single GdTiO$_3$ (GTO)/STO interface, self-consistent Poisson-Schr{\o}dinger calculations show that the interface charge density tail extends $\approx$3 nm from the interface, but that the vast majority of those carriers are located within the first nanometer \cite{Moetakef2011Electrostatic}. This spatial charge distribution should be nearly identical to a SmTO/STO interface, given the similarity between the band structures of GTO and SmTO and their conduction band offsets with STO \cite{bjaalie2016band}. However, when SmTO/STO and STO/SmTO interfaces are separated by less than 2 nm, as is the case for all superlattice films studied here, then the interfacial charge densities overlap significantly and can no longer be considered as isolated 2DES. 

For the 1 SrO (16:1) superlattice, only a single homogeneous slab of Ti in the STO layer makes physical sense, and was consequently the only model considered. On the other hand, several charge distribution models were examined for the 5 SrO (10:5) superlattice. As shown in the Supporting Information, three-slab models allowing for an inhomogeneous charge distribution refine a profile with more carriers at the interface than the interior of the STO layer, exactly as expected from the theoretical calculations \cite{SuppInfo}. However, there is only minimal improvement in goodness-of-fit for these inhomogeneous charge models  ($<$1\%), and therefore we cannot convincingly conclude that they represent a truer picture of the 5 SrO well charge distribution than the homogeneous profile presented in Fig. \ref{fig:SmTORXR_6}(d). This result is in line with previous work on LaTiO$_3$ (LTO)/STO using STEM-EELS that observed only a smooth electron density for heterostructures with thin STO quantum wells embedded within an LTO matrix \cite{ohtomo2002artificial}.

To demonstrate that the apparent homogeneity of the 5 SrO quantum wells is not due to limited spatial resolution, we highlight our sensitivity to the thin Ti$^{4-\delta}$ layers in the limit of 1 SrO thick layers. This is done by creating a model otherwise identical to Fig. \ref{fig:SmTORXR_6}(a), but containing a homogeneous Ti profile throughout the sample below the surface oxidation. The design of this alternative model allows contrast between the SmTO and STO layers generated from the Sm and Sr profiles to be isolated from contrast coming from the Ti$^{3+}$/ Ti$^{4-\delta}$ profiles, by eliminating the difference between Ti valence states. Mathematically, this is accomplished by leaving the layer thickness profile as refined and plotted in  Fig. \ref{fig:SmTORXR_6}(a), but constraining the Ti scattering factors in the STO layers to match those in the SmTO layers. The precise numerical values of those scattering factors were allowed to adjust from their previously refined values to enable the program to attempt to refit the data, and as expected from the small volume fraction of the STO layers, the refined Ti $f^{\prime}$$^{\prime}$ spectrum in this alternative model are essentially identical to the spectrum for isolated SmTO layers.

A pair of 300 K reflectivity fits from this alternative model are plotted in Fig. \ref{fig:SmTORXR_6}(b) for the e$_g$ peak energies (i.e. 460 eV and 465 eV). From this direct comparison, it can be seen that there superlattice peaks in the e$_g$ reflectivity scans (i.e. 460 eV and 465 eV) that are missed when only the Sm/Sr layer contrast is considered (blue). On the other hand, when the Ti$^{3+}$/ Ti$^{4-\delta}$ contrast is included, those superlattice peaks are captured in the fits (red). This can be understood by considering the very large difference in refined $f^{\prime}$$^{\prime}$ values at the e$_g$ peak positions between the SmTO Ti$^{3+}$ and STO Ti$^{4-\delta}$ spectra. This difference is on the order of 50 e/atom, whereas the difference between Sm and Sr is roughly 6 e/atom at the Ti L-edges. It is this large $\Delta$$f^{\prime}$$^{\prime}$ that generates the contrast needed to form a second periodicity in the reflectivity pattern and create scattered intensity at the superlattice peaks. Similarly, the lack of prominent superlattice peaks in the t$_{2g}$ reflectivity curves (i.e. 458 eV and 463 eV), indicates that the contrast between the SmTO Ti$^{3+}$ and STO Ti$^{4-\delta}$ spectra must be small for the t$_{2g}$ orbitals. This is a direct effect of the interfacial charge transfer, and leads to the t$_{2g}$ peak suppression observed in the refined STO $f^{\prime}$$^{\prime}$ spectra.

The comparison in Fig. Fig. \ref{fig:SmTORXR_6}(b) is important for three main reasons. First, it allows us to unambiguously separate the chemical layer profile from the Ti valence profile and isolate the charge transfer in this system. Second, in separating those contrast mechanisms, we confirm that unique Ti valence states are required to capture the reflectivity data. Third, the identification of unique valence state and electronic reconstruction in a 1 SrO quantum well demonstrates resonant X-ray reflectivity is capable of resolving charge transfer at buried interfaces with atomic layer resolution, placing RXR as a non-destructive alternative on a short list of techniques capable of extracting the same information.

In summary, we have used linearly-polarized RXR to measure valence depth profiles for SmTO/STO quantum well heterostructures and probe charge transfer at their buried interfaces. All three samples showed evidence of surface oxidation and a consistent surface termination pattern indicative of SmTO growth dynamics or the stability of SmO-terminated films. The Ti$^{3+}$ scattering factor spectrum of unoxidized SmTO was determined by fitting the depth-dependent RXR data of a 20 nm SmTO control film, and subsequent refinements of the superlattice samples found matching spectra for the buried SmTO layers. The refined $f^{\prime}$$^{\prime}$ spectra of the STO quantum wells revealed a Ti$^{4-\delta}$ valence state with fractional occupation of the t$_{2g}$ levels arising from interfacial charge transfer. This t$_{2g}$ peak suppression was also seen directly in the RXR data through analysis of superlattice peaks and layer contrast, providing an intuitive understanding of the fit $f^{\prime}$$^{\prime}$ spectra. Finally, using a comparative 1 SrO superlattice model without Ti contrast, we confirmed presence of unique Ti valence states in the SmTO and STO layers, and demonstrated RXR's sensitivity to charge transfer at the monolayer limit. 

\acknowledgments{The authors thank Dr. Sebastian Macke for his assistance with the ReMagX refinement software. S.W., R.N., and S.S. acknowledge support under ARO award number W911NF1410379. R.N. was also supported by the National Science Foundation Graduate Research Fellowship under Grant No. 1144085. The Advanced Light Source is supported by the Director, Office of Science, Office of Basic Energy Sciences, of the U.S. Department of Energy under Contract No. DE-AC02-05CH11231.}

\bibliography{SmTORXR_BIBshort}

\end{document}